# Distributed FD-MIMO: Cellular Evolution for 5G and Beyond

*Yeqing Hu, Samsung Research America <yeqing.h@partner.samsung.com>*
*Boon Loong Ng, Samsung Research America <b.ng@samsung.com>*
*Young-Han Nam, Samsung Research America <younghan.n@samsung.com>*
*Jin Yuan, Samsung Research America <j.yuan@samsung.com>*
*Gary Xu, Samsung Research America <gary.xu@samsung.com>*
*Ji-Yun Seol, Samsung Electronics Co., Ltd. <jiyun.seol@samsung.com>*
*Jianzhong (Charlie) Zhang, Samsung Research America <jianzhong.z@samsung.com>*

**Abstract**

This paper presents the next evolution of FD-MIMO technology for beyond 5G, where antennas of the FD-MIMO system are placed in a distributed manner throughout the cell in a multi-cell deployment scenario. This system, referred to as *Distributed FD-MIMO* (D-FD-MIMO) system, is capable of providing higher cell average throughput as well as more uniform user experience compared to the conventional FD-MIMO system. System level simulations are performed to evaluate performance. Our results show that the proposed D-FD-MIMO system achieves 1.4-2 times cell average throughput gain compared to the FD-MIMO system. The insights of performance gain are provided. Hardware implementation challenges and potential standards impact are also presented.

## 1. INTRODUCTION

The cellular industry will see a drastic growth in the wireless data traffic and emergence of new services in the next few years. The amount of data handled by wireless networks will exceed 500 exabytes by 2020 [1]. The 5G cellular system is expected to meet this demand by significantly improving certain key performance indicators, including spectral efficiency, user experienced data rate, peak data rate, areal traffic capacity, network energy efficiency, connection density, latency, and mobility. To bring 5G visions to commercialization, New Radio (NR) standardization effort is well underway in the 3GPP (3rd Generation Partnership Project), where cellular technologies for millimeter wave bands will be introduced, and the fundamental aspects of cellular systems are being redesigned, including waveforms, numerologies, channel coding, and MIMO schemes. 5G NR MIMO schemes are set to encompass key 3GPP LTE MIMO technologies including single-user MIMO, Full Dimension MIMO (FD-MIMO), and Coordinated Multi-Point (CoMP) Transmission [2].

FD-MIMO is the state-of-the-art MIMO technology for 3GPP LTE. The system features a 2D planar antenna array at the base station (BS), which enables tens of active antenna elements to be arranged in a feasible form factor, at operating carrier frequency below 6GHz [3]. FD-MIMO provides significantly higher system throughput and improved user experience than the LTE MIMO systems prior to Rel-12.

However, to satisfy the data traffic requirements beyond 5G, the performance of FD-MIMO requires further improvements. We propose a new FD-MIMO system by spatially distributing the antenna elements of the BS throughout the cell. We refer to such a system as Distributed FD-MIMO (D-FD-MIMO). In particular, D-FD-MIMO is a multi-cell system, where each cell contains tens of distributed antenna elements performing multi-user MIMO

(MU-MIMO). As will be shown later, D-FD-MIMO achieves not only higher system throughput, but also more uniform user experience compared to FD-MIMO.

In this article, we present our vision of D-FD-MIMO as a technology for beyond 5G with the following contributions. First, we provide system level simulation results on the D-FD-MIMO network performance, which conforms to the 3GPP evaluation methodology. A fair comparison between the FD-MIMO and D-FD-MIMO has been conducted. It is shown that D-FD-MIMO outperforms FD-MIMO in certain scenarios. Second, this article demonstrates our D-FD-MIMO experimental results, and discusses practical system implementation aspects. The connection between the distributed antenna elements and the central processing unit imposes challenging hardware requirements. Potential solutions to the raised issues are identified. Third, we discuss potential impacts on the 3GPP standards.

## 2. D-FD-MIMO SYSTEM CONCEPT

D-FD-MIMO is an evolution of FD-MIMO. A D-FD-MIMO network assumes a cellular structure, where a cell is served by one BS and each BS is connected with a large number of antenna elements, of which individual elements are spatially distributed in the cell. One or more antenna elements are equipped with a digital port, and the signals transmitted and received from all the antenna elements within one cell are jointly processed to perform high order MU-MIMO operation.

Such a cellular system can be deployed outdoors in a city-wide area to provide service to both outdoor and indoor users. It can also be deployed inside the building to serve indoor users only. It is also suitable for providing service in a highly populated area, such as stadiums, shopping centers and airports, where a large number of the users are densely located.

Concepts relating to D-FD-MIMO includes distributed massive MIMO, CoMP (a.k.a. network MIMO) and distributed antenna systems (DAS). Distributed massive MIMO [5] treats the entire network as one cell, featuring an enormous number of access points distributed over a large area, jointly serving all the users. pCell [6] by Artemis can be seen as an implementation of the distributed massive MIMO albeit with a smaller scale in terms of the number of antennas. CoMP relies on the coordination among *a few transmission points* from the same or different *sites* to enhance User Equipment (UE) experience at the cell edge [4]. DAS is initially proposed to improve coverage in an indoor cellular communication system, and is sometimes adopted in outdoor scenarios as well. One configuration for outdoor deployment is to have *a few antenna arrays* distributed throughout the *cell* to perform MIMO operations [7]. Another DAS configuration deploys a number of individual antenna elements in a distributed manner in each cell of the network [8], which is similar to the D-FD-MIMO setting. Different from our system-level simulation approach, the analysis in [8] theoretically derives the asymptotic sum capacity when the numbers of UE and antennas in each cell both approach infinity with their ratio fixed, and assuming perfect uplink power control.

## 3. SYSTEM PERFORMANCE EVALUATION

This section presents the system-level simulation results on the downlink performance of the proposed D-FD-MIMO. A comparison between D-FD-MIMO and conventional FD-MIMO is conducted. The network layout contains

19 cell sites in a hexagonal grid, each site serving a fixed number of co-channel UEs. The UEs are randomly located outdoors. The evaluation is performed under large scale channels with the path loss computed according to 3GPP channel models, and the phase computed according to line-of-sight (LoS) condition. No small-scale fading or multi-path is considered. Despite the simplified channel models, valuable insights can be obtained. We consider a carrier frequency of 3.5 GHz, which is a new candidate frequency in NR. Ideal channel state information (CSI) is assumed available at the BS, which is a reasonable assumption for a Time Division Duplex (TDD) system.

### A. EVALUATION SCENARIOS

The conventional FD-MIMO is assumed as the comparison baseline. The system level simulation is performed with 19 cell sites and 57 sectors (i.e., 3 sectors per cell site) deployed in a hexagonal grid. As shown in Figure 1 (a), three 2D antenna arrays are deployed to serve one three-sector site. One antenna array is composed of 32 antenna elements, with 8H x 4V elements and ($0.5\lambda$, $2\lambda$) spacing in horizontal and vertical dimensions, respectively. Each element has a front-to-back ratio of 30dB, peak antenna gain of 8dB, and the half power beamwidth of 65° in both horizontal and vertical directions.

In the D-FD-MIMO evaluation, we consider two scenarios, where one site consists of three sectors or one sector, respectively (as shown in Figure 1 (c) and Figure 1 (d)). For fair comparison with the FD-MIMO reference, one sector contains 32 or 96 distributed antenna elements, in the three-sector-per-site or one-sector-per-site D-FD-MIMO scenario, respectively. In each sector, the omni-directional antenna elements are randomly distributed with a minimum distance of 2 m, and a protection margin of 10 m from the sector edge which helps to reduce the inter-cell interference. The antenna elements in the same sector cooperate to perform MU-MIMO.

As a reference, we also show the results for the D-FD-MIMO deployed as distributed circular arrays, as illustrated in Figure 1(b). We consider a three-sector-per-site layout, where each sector contains 4 circular arrays, each consisting of 8 omni-directional antenna elements. The spacing between two adjacent elements along the circumference is $0.5\lambda$.

The system performance is evaluated for the scenarios described above. We consider the intersite distances (ISD) of 200m and 500m. An exclusion region, where no UEs can reside, is imposed around each antenna, as defined by the 3GPP channel model. Denoting the region of a sector in a three-sector site as an *area*, the performance is evaluated with various numbers of UEs per area. SLNR (Signal-to-Leakage and Noise Ratio) precoding is employed by both the FD-MIMO and D-FD-MIMO, to serve all the UEs in the sector simultaneously with full bandwidth. Figure 1 summarizes the simulation parameters.

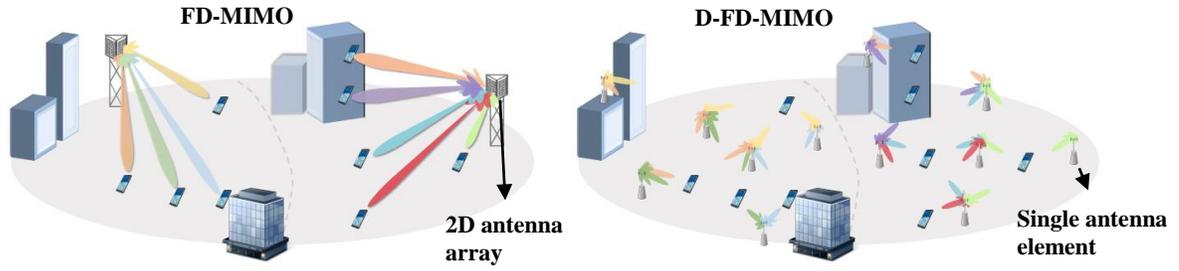

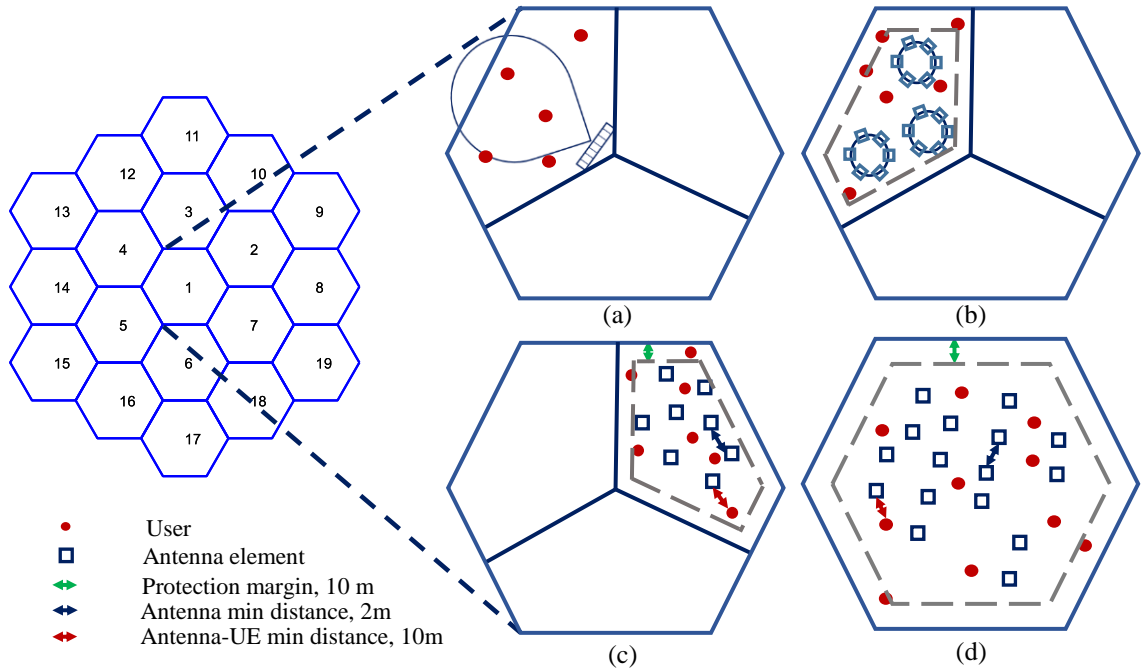

| Scenario | FD-MIMO | Distributed circular arrays (3sec/site) | D-FD-MIMO (3sec/site) | D-FD-MIMO (1sec/site) |
|---|---|---|---|---|
| Element Pattern | 3dB beamwidth (65°, 65°), FBR = 30dB, Gain = 8dB | Omni-directional | Omni-directional | Omni-directional |
| elements/BS | (8H,4V) | 32 (8 elements per circular array, total of 4 arrays) | 32 | 96 |
| Element Spacing | (0.5λ, 2λ) | 0.5λ | ≥2m | ≥2m |
| Protection Margin | NA | 10m | 10m | 10m |
| ISD | 200m, 500m | 200m, 500m | 200m, 500m | 200m, 500m |
| Channel Model | UMi (200 ISD) UMa (500 ISD) | UMi | UMi | UMi |
| Antenna Height | 10m (200 ISD) 25m (500 ISD) | 10m | 10m | 10m |

| Power per BS | 44dBm (200 ISD) | 44dBm (200 ISD) | 44dBm (200 ISD) | 48.8dBm (200 ISD) |
| --- | --- | --- | --- | --- |
| | 49dBm (500 ISD) | 49dBm (500 ISD) | 49dBm (500 ISD) | 53.8dBm (500 ISD) |
| Common parameters | Carrier Frequency = 3.5GHz; Bandwidth = 20MHz; Exclusion Region = 10m; UE Noise Figure = 9dB; UE height = 1.5m; UE/area = {8,16,24,32}; Precoding Scheme = SLNR. | | | |

**Figure 1. Evaluation Scenarios. (a) FD-MIMO with 4x8 antenna elements, (0.5λ, 2λ) spacing. (b) system deploying circular arrays. (c) D-FD-MIMO in a 3-sector-per-site scenario. (d) D-FD-MIMO in a 1-sector-per-site scenario.**

## B. SIMULATION RESULTS

The signal-to-interference-noise ratio (SINR) cumulative distribution function (CDF) and the throughput are computed. Figure 2(a) and Figure 2(b) summarize the area throughput and the 5%-tile UE throughput for ISD = 200m and 500m, respectively. It can be observed that when the ISD is smaller, FD-MIMO yields better cell average throughput than the D-FD-MIMO. However, when the ISD is larger, both scenarios of D-FD-MIMO achieve considerably higher cell average throughput over the FD-MIMO, yielding approximately 1.4 and 2 times gain, respectively. Interestingly, the D-FD-MIMO with a larger ISD yields approximately twice the cell-average throughput compared to the D-FD-MIMO with a smaller ISD, though the 5%-tile UE throughput slightly decreases. On the other hand, such an increase in performance is not observed in the FD-MIMO. In addition, the D-FD-MIMO provides better 5%-tile UE throughput than the FD-MIMO. The one-sector-per-site D-FD-MIMO achieves better performance than the three-sector-per-site D-FD-MIMO, even with the same antenna-to-UE number ratio, indicating that a larger cooperation antenna cluster is beneficial for D-FD-MIMO. In contrast, the circular array deployment does not provide cell average throughput gain over the D-FD-MIMO, and severely degrades the 5%-tile user throughput.

Figure 2(c) and Figure 2(d) show the SINR CDF under the two different ISDs, with 24 UEs located in each area, corresponding to a heavily loaded system. The behaviors described above can also be observed. Under UMi (ISD = 200m) and UMa (ISD = 500m) deployment scenarios, FD-MIMO yields similar performances. On the other hand, when the ISD is larger, D-FD-MIMO SINR shows 8 dB gain in the median, yielding a better overall performance. In contrast, the distributed circular arrays yield better 95%-tile SINR performance but significantly worse 5 percentile SINR performance when the ISD is larger. Although Figure 2(a) shows that the circular array achieves similar performance as the D-FD-MIMO in terms of cell average throughput for ISD = 200m, the SINR CDF indicates that the D-FD-MIMO provides a more uniform UE experience. We also simulate the performance of an isolated cell from the ISD = 200m deployment scenario, as shown in Figure 2(e). In an isolated cell, the D-FD-MIMO achieves remarkably excellent and uniform user experience. The performance of circular arrays lies in between FD-MIMO and D-FD-MIMO. It can be inferred that the inter-cell interference constitutes the major limiting factor for D-FD-MIMO system performance.

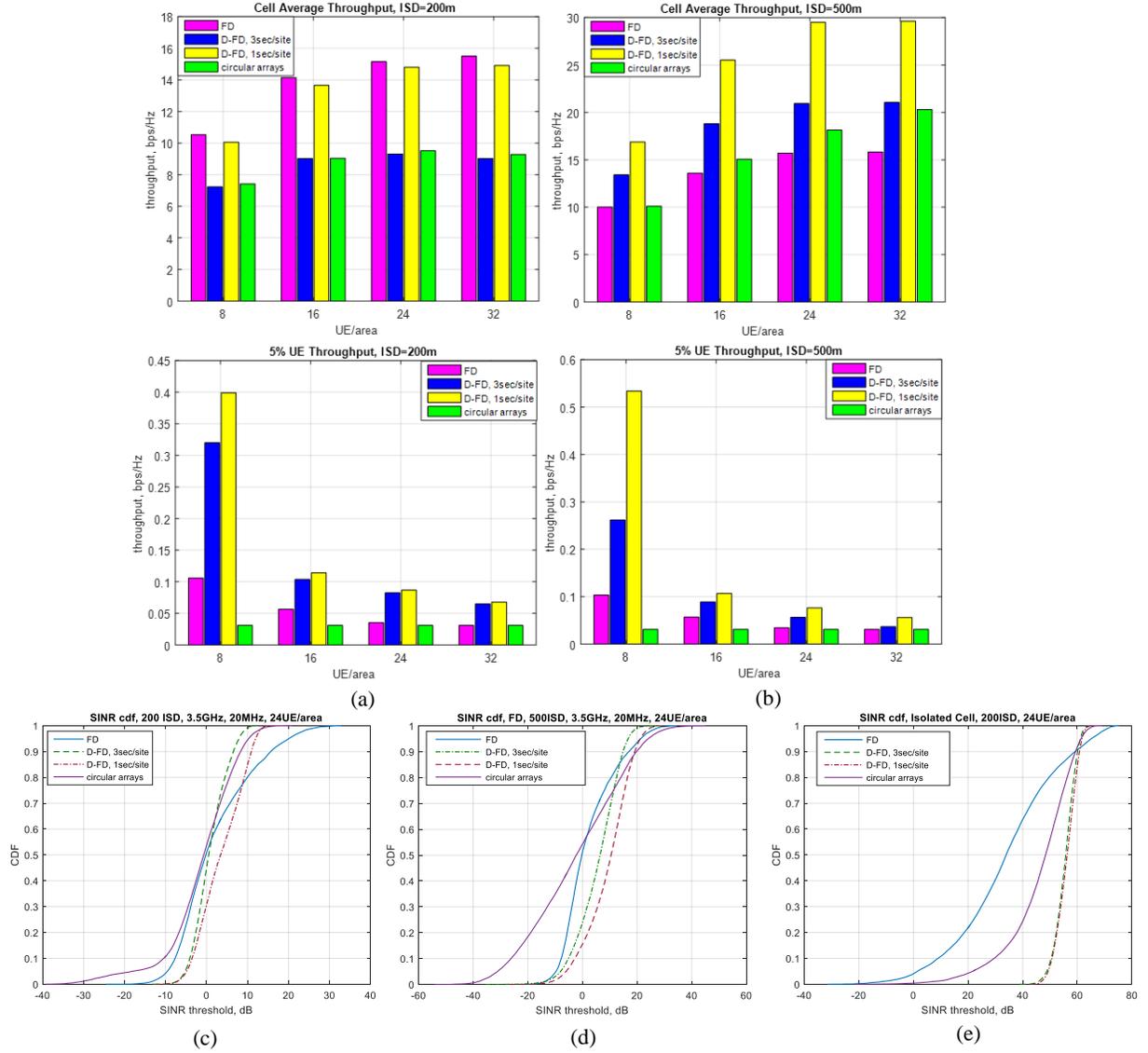

**Figure 2. System performance evaluation results**

In summary, the evaluation results indicate that D-FD-MIMO outperforms FD-MIMO when the ISD is large, or in an isolated cell. Larger cooperation cluster is also more beneficial for D-FD-MIMO performance.

## C. INSIGHTS ON THE PERFORMANCE GAIN OF D-FD-MIMO

This subsection discusses the reasons behind the observations made about the three-sector-per-site D-FD-MIMO and FD-MIMO. Figure 3(a) and Figure 3(b) illustrate the spatial correlation in a single sector, for the D-FD-MIMO and FD-MIMO scenarios, respectively. The magenta dot is the assumed target UE location. The color scale represents the received power intensity (in dB) at each location when the antennas perform conjugate beamforming to the target UE. Since the channel is computed under LoS condition, the power intensity indicates the spatial correlation. For the D-FD-MIMO scenario in Figure 3(a), spatial correlation is generally weak with small variance throughout the cell.

For the FD-MIMO scenario, on the other hand, a wide area of high correlation can be observed. This explains the observation in Figure 2(e) where the D-FD-MIMO yields a steeper SINR CDF curve, due to the fact that it can mitigate the intra-cell interference more effectively.

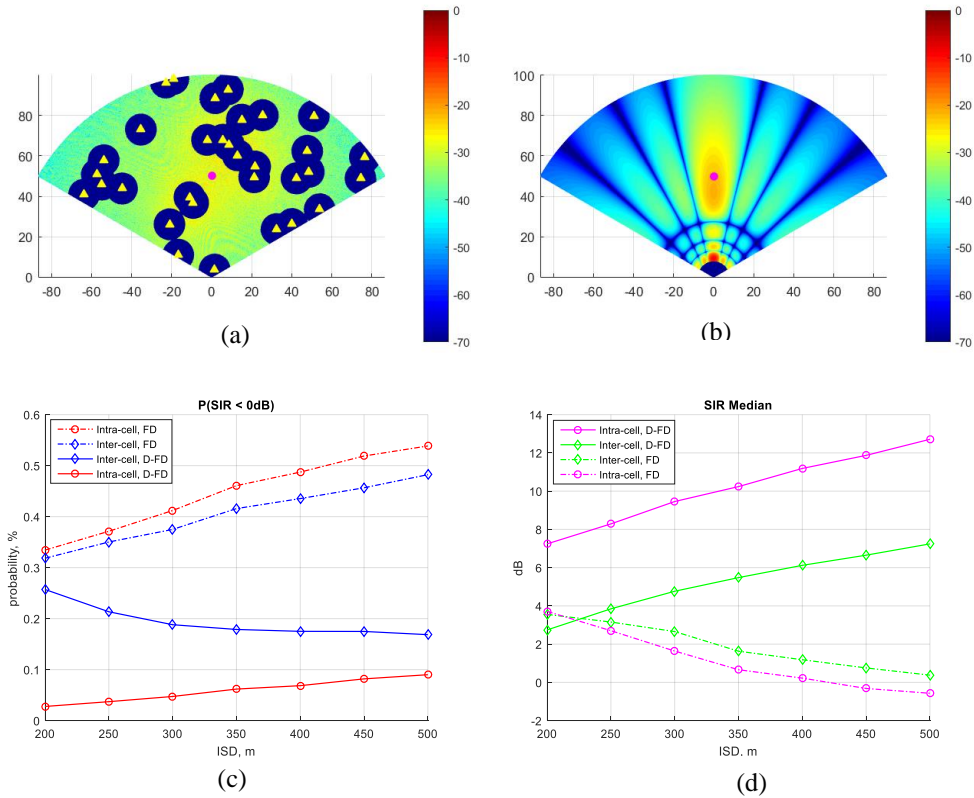

Figure 3. D-FD-MIMO gain insights. (a) Spatial correlation in D-FD-MIMO, yellow triangles denote the 32 randomly distributed antenna elements. (b) Spatial correlation in FD-MIMO. In (a) and (b) magenta dot denoting target UE and dark circles denote exclusion regions. (c) and (d) illustrate the impact of ISD to intra-cell and inter-cell interference

Figure 3(c) and Figure 3(d) show the effects of intra-cell and inter-cell interference, and the impact of ISD. In particular, the signal-to-intra/inter-cell-interference ratios are computed with respect to ISD assuming 24 UEs in each cell. In order to observe the impact of increasing ISDs on these metrics, the transmitting power per sector is fixed at 44dBm and the UMi channel model is assumed for all ISDs. Figure 3(c) shows the probability that the received signal is weaker than the inter- or intra-cell interference. As the ISD increases, these two probabilities increase in the FD-MIMO system. The probability that the received signal is weaker than the intra-cell interference in the D-FD-MIMO system also increases. However, the probability that the received signal is weaker than the inter-cell interference in the D-FD-MIMO system decreases. Hence, it can be concluded that larger ISD attenuates the effect of inter-cell interference in the D-FD-MIMO system. Figure 3(d) shows the median of the signal-to-intra/inter-cell-interference ratio varying with ISD. As the ISD increases, these two ratios decrease in the FD-MIMO system, while increase in the D-FD-MIMO system. Another interesting observation is that the inter-cell interference is more dominant than the intra-cell interference for D-FD-MIMO, while the reverse is true for FD-MIMO. Moreover, the absolute difference between the two kinds of interference is significantly larger for D-FD-MIMO compared to that for FD-MIMO. This reveals that for D-FD-MIMO in a multi-cell setting, mitigation of the inter-cell interference is an important system design goal to realize the full system performance potential of D-FD-MIMO.

## 4. HARDWARE IMPLEMENTATION CONSIDERATIONS

To realize D-FD-MIMO performance gain described in the previous section, one of the main hardware challenges over FD-MIMO is to distribute antennas throughout the cell, which results in the separation of a centralized unit (CU) and multiple remote units (RU). Similar system requirements on hardware in FD-MIMO have to be met in D-FD-MIMO. These requirements include: (1) All ADCs/DACs and RF transceivers need to have coherent clocks; (2) Both DL and UL signals are carried on the links between CU and RUs; (3) CU sends TDD and other control signals to all RUs. Power supply for RUs is another challenge for the D-FD-MIMO system. Normally two options are possible, namely local power supply or remote power supply through cables.

The links between CU and RUs can be seen as the front haul technology in wireless communications. We consider two main front haul technologies, coaxial cable and optical fiber, applicable to D-FD-MIMO and compare them in Table 1. The optical fiber can be further categorized into digital and analogue links. Wireless-based front haul is not taken into account due to the requirements on real time processing and high bandwidth, and the challenges in deployment.

**Table 1: Comparison of coaxial cable link, analog optical fiber link and digital optical fiber link**

|  | Coaxial Cable Link | Analog Optical Fiber Link | Digital Optical Fiber Link |
| --- | --- | --- | --- |
| Max. Length of Link | Short (e.g. 50m @ 3.5GHz, 8dB loss for 1/2" cable). | Long: > 2km. | Long: > 2km. |
| Partitioning Point | Before PA (TX) and after LNA (RX) | Before PA (TX) and after LNA (RX) | Before DAC (TX) and after ADC (RX) |
| Power Supply for RU | Local power or remote through coaxial cable. | Local power supply only. | Local power supply only. |
| TDD and Control Info. | Narrow band signal with analog signal overlay | Narrow band signal with analog signal overlay | Supported by the protocols. |
| Implementation Complexity | Low | Medium | High |
| Cost | Cable cost is high | Medium | High |

The maximum supported link distance is short when the links between CU and RUs are coaxial cable due to the fact that the cable loss is proportional to the cable length. RU is equipped with PA and LNA to mitigate the impact of the cable loss. TDD and control signal can be transmitted to RUs through the coaxial cable by analog signal overlay, in which the control signal is modulated to a narrow band signal and carried at a low frequency (e.g. a FSK signal at 50MHz). In the meantime, the coaxial cable can serve as the media for the power supply of RU, which is a significant benefit to simply the deployment.

Similar to the coaxial cable, the analog optical fiber link transfers RF signal from CU to RUs through optical fiber link, which is called RF over fiber technology (RFoF) [9]. It can support much longer distance which could be up to several kilometers. RFoF technology applies the photonic transceivers based on IM/DD modulation/demodulation schemes, which does not need the clock recovery at the demodulation side. Therefore, no extra frequency shift in the recovered RF signal is induced by the two transducers. Furthermore, the delay and phase rotation caused by the RFoF link is stable after each power on. Therefore, RFoF technology is a perfect candidate for the D-FD-MIMO system with the zero frequency shift and constant delay and phase rotation characteristics. In this setup, RU can be equipped with PA and LNA to compensate the power loss introduced by the pair of transducers at the two ends of the optical fiber link. The similar technology of using narrow band side-band signal as in the coaxial cable link can be used for TDD and control information transfer for RFoF link.

The digital optical fiber link is based on Common Public Radio Interface (CPRI) or Open Base Station Architecture Initiative (OBSAI), which are two sets of protocols defining a flexible interface between CU and RUs. The optical fiber link based on the CPRI or OBSAI standards transfer framed I/Q sample data between CU and RUs, as well as the synchronization, and other user defined information. As a result, ADC/DAC and entire RF transceiver circuits are located in the RU, a major difference in partitioning between CU and RU compared to the analog (RFoF) optical links. The total cost is high [10] and the implementation complexity, at least for RU, is much higher for the digital optical link.

Antenna calibration is a critical process to achieve the performance gain of D-FD-MIMO system running in TDD mode, in which the channel reciprocity is exploited. The purpose of the calibration is to compensate the different response of the transmitter and receiver circuit components. By deploying whole or part of the RF circuit on RU, the integrated calibration using the conventional method [11], is not feasible for D-FD-MIMO system, due to large separation of multiple RUs. One solution is through over-the-air (OTA) calibration, in which the calibration signal is transmitted and received wirelessly by a target transceiver [12]. In a D-FD-MIMO system, each RU antenna can serve as the target antenna in OTA calibration due to the physically distributed RF circuits (especially PAs and LNAs). To deal with NLoS (NLoS between BS antennas) scenario, calibration operation can be achieved by a group based OTA calibration strategy, in which antennas are grouped with at least one overlapping antenna, and the calibration are performed group by group. The group based OTA calibration is shown in Figure 4(a), where the distributed antennas are grouped into three clusters with overlapping antennas. In FD-MIMO, because of the compact array size, the purpose of power calibration is to balance the output power of each antenna element. Keeping the balance of the output power is not important in the D-FD-MIMO case due to different path losses in the distributed antenna array system.

We developed a D-FD-MIMO testbed by using existing FD-MIMO hardware to verify the concept of D-FD-MIMO, and obtained the initial test results for a performance comparison to the FD-MIMO system. The experimental setup consists of a FD-MIMO base station with 32 independent RF chains, of which one half of the 32 RF chains still connect to FD-MIMO collocated antennas, and the other half are connected to 16 distributed antennas through coaxial cables with length of ten meters each. Six TDD UE emulators are connected to the system and the aggregate downlink throughput is monitored as the main system performance metric. After the antenna calibration, we switched between FD-MIMO and D-FD-MIMO by turning on/off the antenna groups dedicated for FD-MIMO and D-FD-MIMO. Figure 4(b) and Figure 4(c) show the system setup, geometric distribution of antennas in D-FD-MIMO and FD-MIMO, and comparison of the test results. The system was first running in D-FD-MIMO and switched to FD-MIMO by turning on FD-MIMO antenna group at 33.8 second. The average aggregate downlink throughput for D-FD-MIMO and FD-MIMO are 150Mbps and 112Mbps, respectively. Similar testing results were recorded when randomly changing the locations of the UE emulators.

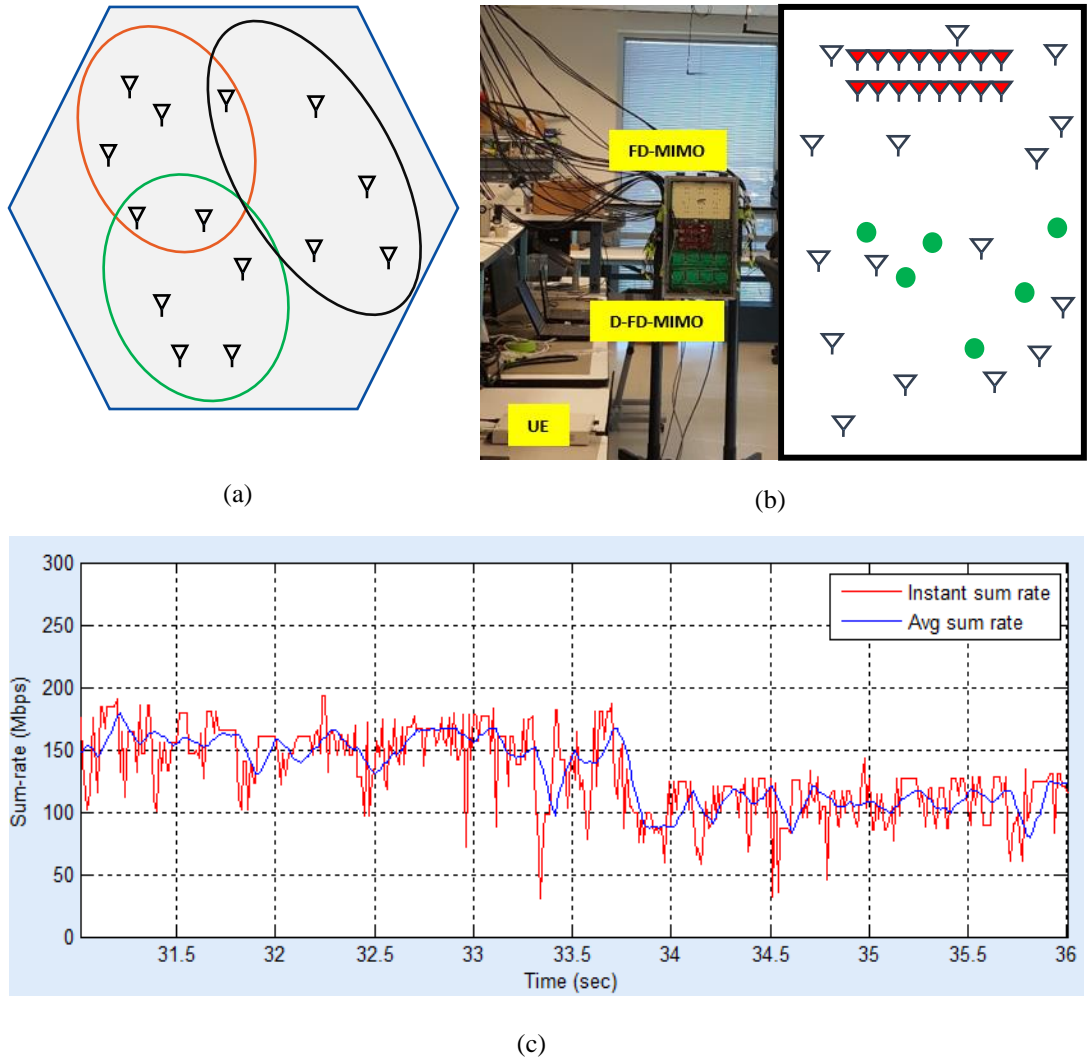

(a)　　　　　　　　　　　　　　(b)

(c)

**Figure 4. (a) OTA calibration: antennas in a cell group into three clusters with overlapping antennas. (b) Laboratory setup and geometrical distribution example of for D-FD-MIMO and FD-MIMO: D-FD-MIMO antennas in white, FD-MIMO antennas in red, and UEs in green. (c) System overall throughput vs time, the switch from D-FD-MIMO to FD-MIMO happened at 33.8 second.**

## 5. FUTURE 5G SPECIFICATION SUPPORT

LTE FD-MIMO was standardized in Rel-13, where up to 16 Channel State Information Reference Signal (CSI-RS) ports can be configured to a UE. This is followed by Enhanced FD-MIMO (eFD-MIMO) in Rel-14, where the maximum number of supported CSI-RS ports is increased to 32. However, both FD-MIMO and eFD-MIMO were designed for co-located antennas. Further MIMO evolution is underway with the on-going standardization of 5G NR. In NR MIMO, a single streamlined framework will be introduced to support the family of key 3GPP MIMO technologies including single-user MIMO, FD-MIMO and CoMP. Advanced CSI feedback that enables measurement and reporting of high resolution spatial information by the UE is also expected to be supported. Nevertheless, D-FD-MIMO represents a new deployment scenario yet to be studied in detail in 3GPP. Hence, further study would be needed on the enablers for supporting D-FD-MIMO under various conditions including a variety of indoor or outdoor environments, cell sizes, UE mobility, antenna densities, and UE densities.

Conventional cellular systems handle L3 mobility based on downlink measurement, e.g., based on cell-specific RS (CRS) in LTE. Support for L1/L2 mobility handling (intra-cell mobility) will be a major enhancement in 5G NR. However, the mobility handling for 5G NR is expected to remain inherently downlink measurement based. D-FD-MIMO requires support of an efficient mechanism to frequent UE-specific antenna set reconfiguration due to UE mobility. Hence, downlink based L1/L2 mobility may need to be enhanced to support low latency measurement and reporting cycle. UE measurement reporting overhead should also be minimized.

To circumvent the latency and UE reporting overhead incurred by the downlink based mobility mechanism, uplink based mobility can be considered. To support uplink based mobility, the UE can be configured to transmit an uplink reference signals (e.g., Sounding RS (SRS) or Physical Random Access Channel (PRACH), either periodically or on-demand), which is to be measured at the antennas of the D-FD-MIMO network. Based on uplink measurement, the D-FD-MIMO network dynamically configures the set of antennas associated with each UE for CSI feedback, thereby avoiding the latency of measurement, antenna selection and reporting by the UE.

A table summarizing comparisons of standards impact between LTE eFD-MIMO, NR MIMO, and D-FD-MIMO is given in Table 2.

**Table 2: Comparisons of standards impact between LTE eFD-MIMO, NR MIMO, and D-FD-MIMO**

|  | LTE eFD-MIMO (Rel-14) | NR MIMO (Rel-15) | D-FD-MIMO |
|---|---|---|---|
| Number of CSI-RS ports | 32 | 32 (expected) | >32 |
| Advanced CSI feedback | Advanced rank 1-2 codebook for MU-MIMO | Higher resolution codebook | Similar resolution as NR MIMO. New codebook design (FDD) |
| L1/L2 mobility | RRC configured a set of CSI-RS resources with dynamic resource trigger (aperiodic & semi-persistent CSI-RS) | Expected to be similar to LTE eFD-MIMO | UL measurement assisted to reduce latency |
| L3 mobility | Based on CRS measurement | Based on measurement of SS block and possibly an additional cell-specific RS | Same as NR MIMO |

## 6. CONCLUSIONS

D-FD-MIMO is the next generation of FD-MIMO system for beyond 5G, which has the potential to increase the cell average throughput of conventional FD-MIMO by 1.4-2 times, while providing more uniform user experience. Due to practical deployment considerations such as hardware implementation of links between the CU and RUs, D-FD-MIMO in a multi-cell setting is an important deployment scenario. Our analyses indicates that inter-cell interference mitigation is one of the key system design goals to further improve the performance of D-FD-MIMO. We presented laboratory experiment setup and result of a D-FD-MIMO system, which illustrates the feasibility of the system and the performance gain over FD-MIMO. We concluded with an initial summary of beyond 5G standardization effort that may be needed to support D-FD-MIMO.